\documentclass[review]{elsarticle}

\usepackage{lineno,hyperref}
\modulolinenumbers[5]
\usepackage[margin=1in]{geometry}

\usepackage{graphicx}
\usepackage{listings}
\usepackage{makecell}
\usepackage[utf8]{inputenc}
\usepackage[T1]{fontenc}

\usepackage{mathpazo} 
\usepackage{caption}
\usepackage{paralist}
\usepackage[justification=centering]{caption}
\usepackage{setspace}
\usepackage{algorithm}

\usepackage{algpseudocode}

\usepackage{sectsty}
\sectionfont{\fontsize{10}{10}\selectfont}

\journal{Computers and Security}

\begin{document}

\begin{frontmatter}
\title{A Comparative Risk Analysis on CyberShip System with \\ STPA-Sec, STRIDE and CORAS}

\author[1]{Rishikesh Sahay}
\ead{rishikesh.sahay@oit.edu}

\author[2]{D.A.Sepulveda Estay}
\ead{daniel.alberto.sepulveda.estay@regionh.dk}

\author[3]{Weizhi Meng}
\ead{weme@dtu.dk}

\author[3]{Christian D. Jensen}
\ead{cdje@dtu.dk}

\author[4]{Michael Bruhn Barfod}
\ead{mbba@dtu.dk}

\address[1]{Business Management Dept., Oregon Institute of Technology, Klamath Falls, 97601, Oregon, United states}
\address[2]{Digitalization Group, Rigshospitalet}
\address[3]{Dept. of Applied mathematics \& Computer Science,Technical University of Denmark, DK-2800 Kgs., Lyngby, Denmark}
\address[4]{DTU Management Engineering,Technical University of Denmark, DK-2800 Kgs., Lyngby, Denmark}

\begin{abstract}

The widespread use of software-intensive cyber systems in critical infrastructures such as ships (CyberShips) has brought huge benefits, yet it has also opened new avenues for cyber attacks to potentially disrupt operations.
Cyber risk assessment plays a vital role in identifying cyber threats and vulnerabilities that can be exploited to compromise cyber systems.
Understanding the nature of cyber threats and their potential risks and impact is essential to improve the security and resilience of cyber systems, and to build systems that are secure by design and better prepared to detect and mitigate cyber attacks.
A number of methodologies have been proposed to carry out these analyses.
This paper evaluates and compares the application of three risk assessment methodologies: system theoretic process analysis (STPA-Sec), STRIDE and CORAS for identifying threats and vulnerabilities in a CyberShip system.
We specifically selected these three methodologies because they identify threats not only at the component level, but also threats or hazards caused due to the interaction between components, resulting in sets of threats identified with each methodology and relevant differences.
Moreover, STPA-Sec which is a variant of the STPA is widely used for safety and security analysis of cyber physical systems (CPS); CORAS offers a framework to perform cyber risk assessment in a top-down approach that aligns with STPA-Sec; and STRIDE (Spoofing, Tampering, Repudiation,Information disclosure, Denial of Service, Elevation of Privilege) considers threat at the component level as well as during the interaction that is similar to STPA-Sec.
As a result of this analysis, this paper highlights the pros and cons of these methodologies, illustrates areas of special applicability, and suggests that their complementary use as threats identified through STRIDE can be used as an input to CORAS and STPA-Sec to make these methods more structured.

\end{abstract}

\begin{keyword}
Cyber ship, Cyber Physical Systems (CPS), STPA, Cyber Risk Assessment, STRIDE, CORAS
\end{keyword}

\end{frontmatter}


\section{Introduction}
\label{sec:intro}

Modern day ship systems are highly sophisticated, complex and dependent on the effectiveness of software-based systems for operation.
Until 15-20 years ago, these ship systems were not connected to the Internet, security was restricted only to safeguarding the physical infrastructure~\cite{CFCS}.
The present internet connectivity in these systems, despite providing ease of operation, yet has also exposed them to cyber threats and vulnerabilities which are difficult to prevent with a strategy based only on physical infrastructure safety.
Additionally, the advent of Industry 4.0 in the maritime industry is advancing the use of process digitalization, and the use of machine learning for data analysis allowing automated decision making and operation.
This dramatically expands the attack surface for cyber attacks on ship systems.

In June 2017, A.P. Moller-Maersk was attacked by a malware known as Not-Petya that left its IT systems inoperable for several weeks~\cite{notpetya}.
The estimated damage due this attack on Maersk is \$300 million~\cite{lost_revenue}.
It caused the disruption in the global supply chain and impacted many other companies along with Maersk.
Many reports suggest that the attack caused as much as \$10 billion in damages in total~\cite{loss_notpetya,notpetya_10}.
Beyond the immediate effects that this attack had on Maersk's bottom line, this was another clear evidence that cyber attacks that can go beyond the loss or corruption of data, to cause operational disruptions, are also a reality in the shipping industry.

Moreover, nowadays shipbuilders are also trying to innovate and build remote controlled automated ship.
For instance, the Nippon foundation launched MEGURI2040 fully autonomous ship program in February 2020.
On January 2022, a fully autonomous small boat successfully sailed in the waters around Sarushima in Japan~\cite{autonomous_boat}.
These autonomous ships are equipped with Operational Technology (OT) systems, that are interconnected with each other to automatically navigate the ship.
According to the Danish Centre for Cyber Security (CFCS), cyber threats against OT systems of ships is high~\cite{CFCS}.
CFCS assess that the OT systems of ships can be attractive target for cyber criminals, since they are important for the shipping company, and they can leverage it for getting ransom.
The CFCS's report highlights that in 2017 a ransomware attack spread through administrative system to the ship's OT systems and disrupting the power supply~\cite{CFCS}.
The crew was not able to solve the issue and had to call help from the IT support~\cite{CFCS}.
Some works have highlighted the cyber security risks on autonomous ships~\cite{cyberrisk_autonomousship,cyberrisk_autonomousvessel}.

Many other less visible attacks are happening to shipping operations every day, in a trend that is showing no signs of slowing down~\cite{CFCS}.
Companies in the shipping industry, formerly inclined to invest mainly in cyber security, have increasing evidence that failing to avert a cyber-attack is more and more likely.
Cyber-resilience, the capacity to react to cyber-attacks, becomes thus desirable, through for example, designing a system with the ability to cope with a cyber attack already under way through DCRA resilience, namely Detection, Contention, Recovery and Adaptability.

In light of these findings, it is vital to improve the cyber security of ship systems.
To strengthen the cyber security of these systems, it is important to have a holistic view of ship systems.
Therefore, the first step is to prepare an architecture and identify System under Consideration (SuC) for cyber risk assessment.
Second step is to perform cyber risk assessment and mitigation.
Moreover, cyber risk assessment should not only consider the risk at the component level but also the way it can propagate to interconnected components and compromise the whole system.
Components in these critical infrastructures such as ships are interconnected, therefore the disruption in one part of the system can trigger domino effect causing the damage to the whole system.
So, having a holistic system-of-systems approach is important in cyber risk analysis of critical infrastructures.

Therefore, this paper compares three different risk assessment methodologies namely STPA, STRIDE and CORAS.
For the purpose of comparison we apply it on the CyberShip framework~\cite{CyberShip}.
Therefore, this paper presents a CyberShip framework comprised of cyber physical components of the ship.
The main purpose of using these methodologies is to investigate how they perform in analyzing cyber and safety risks on critical infrastructure mainly CyberShip systems in this paper.
Moreover, all three methodologies follow top down approach in analyzing risks.
Furthermore, they also consider the risk because of interaction between the different components of the system.

The rest of the paper is organized as follows.
Related extant literature on threat modelling through the use of systemic risk analyses is described in Section~\ref{sec:rel_works}.
Section~\ref{sec:cybership_framework} describes the CyberShip framework and its different components.
Thereafter this CyberShip framework is analyzed using STPA-Sec in Section~\ref{sec:cybership_stpa}, by using the STRIDE method in Section~\ref{sec:stride_cybership}, and through the use of the CORAS method in Section~\ref{sec:coras_cybership}.
Thereafter these analyses are compared in Section~\ref{sec:risk_analyses_comparison}.
Section~\ref{sec:discussion} presents a discussion of the comparison of these risk analyses, and finally, Section~\ref{sec:conclusion} concludes the paper proposing areas of future work.

\section{Related works}
\label{sec:rel_works}

A number of guidelines have been developed to address the growing concern of cyber attacks in the maritime industry~\cite{bimco_guidelines,dnvgl,abs}.
These guidelines provide a framework for securing ship systems and its operations.
Moreover, a number of studies have been done on the cyber risk assessment of the ship systems and networks.
MaCRA (Maritime Cyber Risk Assessment) model developed by Tam and Jones~\cite{tam2019macra} dynamically responds to the changes done within ship system and to threats.
It considers the vulnerabilities in the system as well as how easily it can be exploited by adversaries.
The authors in~\cite{cyberrisk_autonomousship} proposed a generic system architecture of autonomous ship and analysed threats and risks using STRIDE model.
The AAWA (Advanced Autonomous Waterborne Applications) project led by Rolls Royce has also done an important work in highlighting cyber security and safety issues in autonomous ships~\cite{rolls-royce}.
System Theoretic Process Analysis (STPA) have been used in analyzing cyber security risks in ship systems.
For example, STPA has been used to derive verification objectives and hazardous scenarios in maritime systems \cite{rokseth2018deriving}, to identify causal scenarios and factors that drive maritime incidents and accidents \cite{puisa2018unravelling} and it has been advanced in the conceptual design autonomous vessels \cite{banda2019systemic}.
STPA has been used to identify the conditions of risk for the case of remotely-controlled merchant vessels \cite{wrobel2018system}, work that has focused mainly on the overall shipping operation, considering the vessel, the shore facilities, the environment and the organizational environment, all in an aggregated level.
Recent research that describes the multiple control systems on board standard commercial ships \cite{bartek}, reflects a need for greater detail in the systemic analysis of risks, a suggestion that is developed in this work.
The authors in~\cite{navigation_secure} used STPA-Sec which is a variant of STPA method to analyse cyber attacks targeting navigation system of ships.
STPA-Sec focuses on unsafe control actions (UCA) which occur due to cyber threats.
In~\cite{stpa-attack} the authors suggested to use the STPA with attack tree for safety and security analysis of autonomous vessels.
Kavallieratos et.al.~\cite{SafeSec} employed STPA method with security analysis to find the comprehensive list of security and safety requirements at the system design stage.
In connection to cyber risks, authors have also proposed the combined use of STRIDE with STPA \cite{kaneko2018threat}.

More recently a group led by Lim~\cite{lim2018models} has identified models and computational algorithms used in maritime risk analysis and mentions that the development of models was the most common type of maritime risk analysis research.
Akpan et.al. \cite{maritime_challenges} highlights the cyber risks in the various OT (Operational Technology) components of the ship.
Grigoriadis et.al.~\cite{adaptive_riskassessment} proposed an adaptive security framework to identify situational risks and deploy policy to mitigate them.
SafeSec~\cite{SafeSec} offers a framework to identify security and safety objectives in the ship systems.
This framework combines model based method for security requirements engineering with STPA method to identify safety risks.
Wang et.al.\cite{wang2004use} described control engineering techniques that could be used for the risk analysis in the marine industry, highlighting that \textit{"a framework with a holistic nature is desirable for risk assessment of large engineering systems"}.

Most of the previous studies have been conducted to identify cyber and safety risks in the ship systems either with one method or by combining two methods.
However, this paper identifies cyber and safety risks with STPA, CORAS, and STRIDE methods.
STPA, CORAS and STRIDE identify risks in a top down approach and offers a structured framework for risk assessment.
The aim is to find out how these methods perform in terms of identifying security and safety risks when applied on the same framework CyberShip in this case.

\section{The CyberShip Model}
\label{sec:cybership_framework}

    CyberShip is a Cyber Physical System \cite{kayan2021cybersecurity} generated from the widespread adoption of Information and Communication Technologies (ICT) in maritime operations, and as a result increasing the attack surface of ships to cyber attacks \cite{bartek}.
    The risk analyses presented in this paper are based on a CyberShip model of a system representing the interaction between infrastructure and information.

    The CyberShip model used in this study is a hierarchical control structure as shown in Figure \ref{fig:cybership_framework}.
    This representation is consistent with literature \cite{chaal2020framework,bartek,cybership2019,cybership2020} and includes several cyber-physical systems that are important for the safe operation of the ship:

    \begin{figure}[h]
        \includegraphics[width=0.9\textwidth]{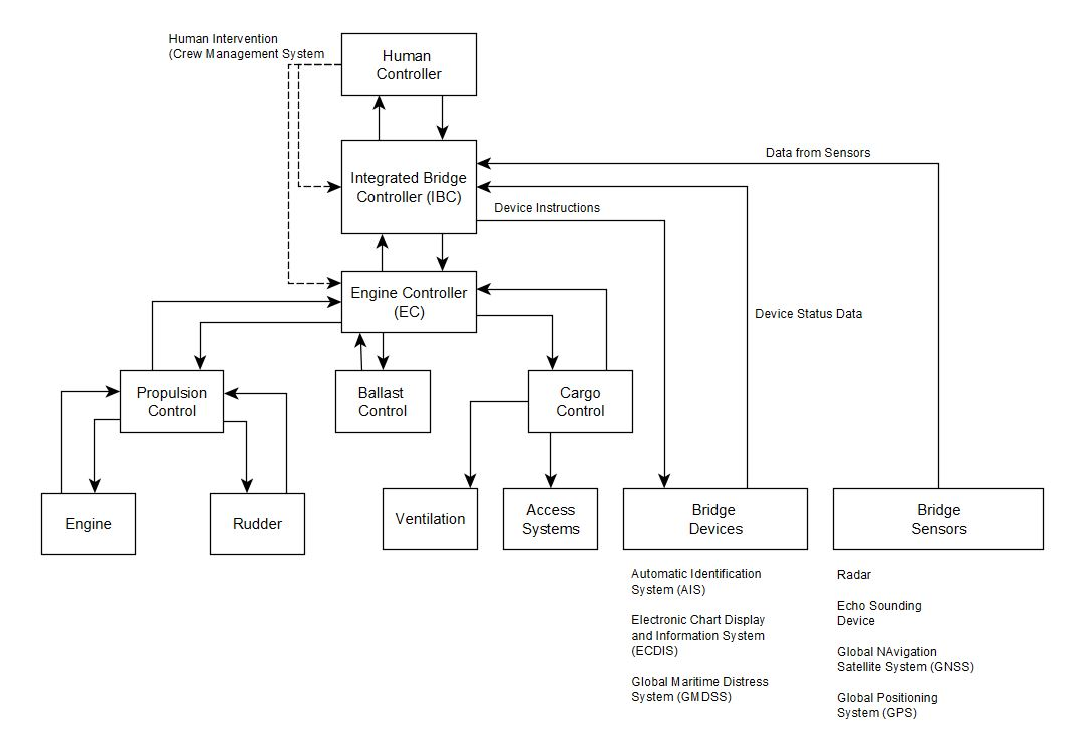}
        \centering
        \caption{CyberShip Framework}
        \label{fig:cybership_framework}
    \end{figure}

    \begin{itemize}

        \item \textbf{Bridge Devices:} These devices sense the surrounding environment and provide this information to the ship controller for centralized process control and decision making.
        Bridge devices can be connected to shore-side networks for software updates, or be updated through removable media such as USB.
        Radar, Automatic Identification System (AIS), Electronic Chart Display System (ECDIS), Global Maritime Distress System (GMDSS), Global Navigation Satellite System (GNSS), and Echo Sounding are examples of bridge devices on the ship~\cite{bimco_guidelines}.

        \item \textbf{Integrated Bridge Controller (IBC):} It supervises the operation of bridge devices \cite{bimco_guidelines} by receiving data from sensors in these devices and providing a centralized interface to the crew on-board to access the data and to make decisions.
        The IBC also issues control commands to the engine controller, such as start/stop of the propulsion control system, rerouting the ship, and increase/decrease water level in the ballast, depending on the information from the bridge devices~\cite{bimco_guidelines}.

        \item \textbf{Engine Controller:} It controls all the system related to power generation and propulsion~\cite{autonomous_engine}.
        It gathers data related to speed, rudder angle, and propeller, and it monitors the engine load, fuel consumption, and water level in the ballast compartment.
        Depending on the information from the integrated bridge controller, the engine controller commands and controls the to propulsion control system to increase or decrease the speed of the ship.
        Furthermore, it also sends the command to increase or decrease the level of water in the ballast compartment depending on the information from the bridge system.

        \item \textbf{Ballast Water Control:} It supervises the operation of the the ballast tank system in the ship used for draft and balance control~\cite{Munin_process}.
        Ballast tanks are compartments within a ship that hold water, and which is used to provide stability, by adjusting the ship balance.
        If the water in the ballast tanks is pumped out temporarily, this reduces the draft of the vessel.
        Depending on the model of the ship, the ballast water control is independent of the engine system.

        \item \textbf{Propulsion Control:} It controls the propeller, rudder and steering of the ship.
        Propulsion control acts on the inputs from the engine control and provides the information to the engine controller such as speed of the ship, fuel level, engine load, etc~\cite{bimco_guidelines}.

        \item \textbf{Cargo Management System:} Computer systems used for the management and control of cargo may interface with a variety of other systems ashore~\cite{bimco_guidelines}.
        These systems may include shipment tracking details available to shippers via the Internet.
        Interfaces of this kind make cargo management systems and data in cargo vulnerable to cyber attacks.

        \item \textbf{Human factors} also have to be considered in a cyber-ship model, as only in highly automated shipping systems there is no expected interaction between human operators and the shipping system.
        Examples of human factors that can have a disruptive effect through cyber-attacks include events such as unauthorized system entry (software level) or rewiring (hardware level).

    \end{itemize}


\section{CyberShip analysis using STPA-Sec}
\label{sec:cybership_stpa}

\begin{figure}[h]
    \includegraphics[width=0.5\textwidth]{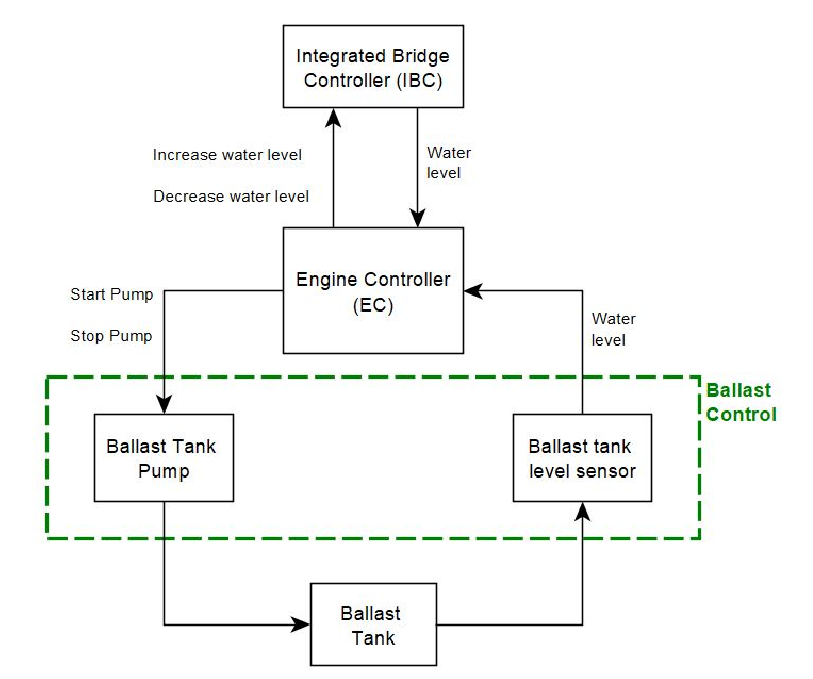}
    \centering
    \caption{High-level Structure of the ballast control}
    \label{fig:control_loop}
\end{figure}

This section first describes the STPA risk analysis methodology to then present the control structure of a \texttt{CyberShip} ballast control system.
A high level control diagram of a \texttt{CyberShip} ballast control system is represented in Figure \ref{fig:control_loop}.
This section closes by identifying undesirable behaviors to be avoided and the control actions required to avoid hazardous situations.

\subsection{STPA: Systems Theoretic Process Analysis}
\label{subsec:stpa_analysis}

The STPA is a risk analysis methodology for safety and security, based on systems theory rather than traditional analytic reduction and reliability theories.
It conceptualizes losses as a result of the inadequate interaction between components in the system due to a lack of adequate safety constraints.
Consequently, safe and secure operation is seen as an emergent property resulting from the interactions between system components and the environment \cite{leveson2011engineering}.

In STPA, events that lead to a system failure are known as \texttt{accidents}, and these can occur through component failure for example, condition that has been extensively analyzed through failure mode analysis methods such as FMEA or HAZOP~\cite{fmea_hazop}.
Other more subtle system failure types are caused by unintended component interactions even when no component failure occurs.

Components are controlled through the enforcement of constraints and STPA assumes that inadequate constraints at different system levels lead to system failures and accidents.
Once a control structure and the interactions among components are represented, STPA suggests a safety and security analysis from a broad perspective including aspects such as physical, logic and information, social, operational, and managerial.

By analyzing the system's control structure, STPA represents how the interaction of different components can result in a safe and secure system, particularly by proposing requirements that prevent unsafe control actions, a result that is not possible with the traditional risk assessment mechanisms.

An STPA-Sec method identifies potentially unsafe control actions by analyzing how these can lead to an accident.
Four ways in which a control action can be unsafe are 1) if a control action is executed, leading to a hazard, 2) if a control action is not executed, leading to a hazard, 3) if a control action is provided too early or too late, leading to a hazard, or 4) if a control action is executed for too long or stopped soon, leading to a hazard \cite{leveson2011engineering}.
After the unsafe control actions have been identified, the next step involves investigating the system control structure to identify conditions which can lead to these unsafe control actions happening.


First, the control loops are identified in the system under analysis.
Second, the hazardous control actions are identified. \
Third, these hazardous control actions are used to define security requirements and constraints. Finally, causal scenarios are identified which lead to a violation of either safety or security constraints.

By representing cyber and safety risks through a control system structure, reflects that cyber-attacks are not events that happen from external sources, but rather events which a system such as CyberShip are \textit{“mis-designed”} to experience.
In this approach, risky cyber-events are \textit{unintended consequences} resulting from incomplete requirements at the time of system design.

A systemic analysis such as STPA-Sec follows a process to identify the “unintended” design that creates cyber-vulnerability, and suggests design changes through which cyber-vulnerable behavior is less likely or no longer possible.

\subsection{Unacceptable losses and hazards}
\label{sec:undesirable_loss}

In the first step, we identify all the losses which are considered unacceptable.
In this stage, we identify what are the essential functions which must be defended and how the disruption of these functions can lead to the undesirable outcomes.
The analysis goes from top to down, from high-level to concrete.
Table~\ref{tabl:loss} shows losses which are unacceptable and must be avoided.
These include service related losses such as late arrival of the shipment, wrong delivery to customers, or damage to the cargo, operational losses such as damage to the ship, human losses, and business losses from damage to the reputation of the shipping organization.

\begin{table}[h]
    \centering
    \caption{Undesirable behavior or loss}
    \label{tabl:loss}
    \begin{tabular}{|l|l|}
        \hline
        A1 & Shipment late or non arriving         \\ \hline
        A2 & Loss/harm to life of passengers/crew  \\ \hline
        A3 & Wrong or non delivery to customers    \\ \hline
        A4 & Damage to the ship                    \\ \hline
        A5 & Damage to the cargo                   \\ \hline
        A6 & Reputational loss                     \\ \hline
    \end{tabular}
\end{table}

Based on the control structure shown in Fig.~\ref{fig:control_loop}, our research lists hazards in Table~\ref{tabl:hazard} which can lead to the loss shown in Table~\ref{tabl:loss}.
Hazard is defined as a condition which can lead to high-level loss \cite{leveson2011engineering}.

\begin{table}[h]
    \centering
    \caption{Hazards}
    \label{tabl:hazard}
    \begin{tabular}{|l|l|}
        \hline
        H1 & Uncontrolled maneuvering of the ship      \\ \hline
        H2 & Unidentified cargo items/wrong cargo data \\ \hline
        H3 & Incorrect functioning of ship components  \\ \hline
        H4 & Uncontrolled transmission of data         \\ \hline
        H5 & Uncontrolled data being transmitted       \\ \hline
    \end{tabular}
\end{table}

The Next step of analysis involves the identification of unsafe control actions (UCAs) which can lead to the hazards as shown in Table \ref{tabl:hazard}.

The CyberShip control structure shown in Fig.~\ref{fig:control_loop} combined with the identified high-level losses and hazards set the foundation for STPA analysis.
We consider the control actions \texttt{start pump} from Engine Controller (EC) to the ballast pump and then identified the UCAs which can cause hazard, in any of the four types outlined by Leveson \cite{leveson2011engineering}

Table~\ref{tabl:UCA} shows the list of unsafe control actions related to the action performed by ballast pump based on the information provided by the Engine controller (EC).
For instance, \texttt{start pump} action leads to increase or decrease of the level of water in the ballast tank.
Increase or decrease of the water to a wrong level in the ballast tank can imbalance and damage the ship.
The Engine controller might have received wrong parameters from the Integrated Bridge Controller which can cause EC to initiate \texttt{start pump} action with the wrong parameters, which can cause hazard.

The \texttt{Start pump} action can cause hazards in different conditions.
For instance, if the ship is sailing through a shallow or deep water, then it may require to increase or decrease the level of water in the ballast tank to balance the ship depending on the scenario.
However, if the \texttt{Engine Controller} is compromised by an external adversary or is not functioning properly because of the component failure then it can damage the ship.

It should be noted that the attacker can modify the parameters instructing the ballast tank to increase or decrease the water to a wrong level which can sink the ship.

This example highlights a main advantage of this methodology that the individual components are working well but the vulnerability lies in the interaction between different components.
Table~\ref{tabl:UCA} highlights example of conditions when the control action CA1 \texttt{start pump} can become unsafe and lead to the undesirable losses shown in Table~\ref{tabl:loss}.

\begin{table}[h]
    \small
    \centering
    \caption{Unsafe Control Actions for Start Pump Action from Engine Controller (EC) to Ballast Tank Pump}
    \label{tabl:UCA}
    \begin{tabular}{|p{1cm}|p{3,25cm}|p{3,25cm}|p{3,25cm}|p{3,25cm}|}
        \hline
        Control Action & Performed with Hazard & Not performed  with hazard & Performed too long or too short with hazard & Performed too early or too late with hazard \\ \hline

        \textbf{CA1}: Start Pump
        &
        \textbf{UCA1.1}: when EC has provided wrong parameter (Velocity, Level) to Pump. \newline
        \textbf{UCA1.2}: when EC receives the wrong parameters from IBC \newline
        \textbf{UCA1.3}: when Ballast tank Pump is not functioning. \newline
        \textbf{UCA1.4}: when Due to network failure control action is not received by Ballast tank. \newline
        \textbf{UCA1.5}: when EC is compromised because of human in the loop. \newline
        \textbf{UCA1.6}: when EC is compromised because of component failure. \newline
        \textbf{UCA1.7}: when EC is compromised because of external hacker \newline
        \textbf{UCA1.8}: when it was not required.
        &
        \textbf{UCA1.9}: when EC is compromised because of human in the loop. \newline
        \textbf{UCA1.10}: when EC is compromised because of component failure. \newline
        \textbf{UCA1.11}: when EC is compromised because of external hacker. \newline
        \textbf{UCA1.12}: when EC did not receive command from IBC.
        &
        \textbf{UCA1.13}: when \texttt{requirement} was for a shorter period and the pump acted for too long. \newline
        \textbf{UCA1.14}: when \texttt{requirement} was for a longer period and the pump acted for too short.
        &
        \textbf{UCA1.15}: when there are communication channel congestion. \newline
        \textbf{UCA1.16}: when there is a feedback delay between Actuator to Ballast tank. \newline
        \textbf{UCA1.17}: when EC action was performed too early or too late.
        \\\hline
    \end{tabular}
\end{table}

\subsection{System Security Constraint and Security Requirements}
\label{subsec:system_security_constraint}

The analysis of unsafe control actions is used to suggest design requirements and constraints.
These suggestions result from the conditions where the control actions become unsafe.

For example the following constraint should be specified: the \texttt{start pump} action must not be provided if the water level information is not received from \texttt{Integrated Bridge Controller}. This responds directly to \textbf{UCA1.4}.
The specific implementation of this constraint is not specified, as it can be achieved in different ways.

In the same way, a requirement (a risk boundary) should be set to avoid the increase or decrease of water level in the ballast tank to a dangerous level.
This boundary will provide protection against commands issued from a compromised \texttt{Integrated Bridge Controller} to \texttt{Engine Controller} with wrong water level information, or avoid damage caused in case the \texttt{start pump} action is applied for too long or too short time.

Following this process, system constraints and requirements can be proposed for scenarios derived from the analysis of the other \texttt{Control Actions} designed in the system. Examples of other constraints and requirements are shown in Table \ref{tabl:requirements-constraints}.

\begin{table}[h]
    \footnotesize
    \centering
    \caption{Requirement and constraint examples}
    \label{tabl:requirements-constraints}
    \begin{tabular}{|p{2cm} p{8cm}|}
        \hline
        \textbf{Constraints} &       \\ \hline
        C1 & {\texttt{start pump} action must not be provided if the water level information is not received from \texttt{Integrated Bridge Controller}}  \\ \hline
        C2 & Parameters communicated for \texttt{action} needed before execution  \\ \hline
        C3 & User interface limited to required \texttt{actions}  \\ \hline
        C4 & \texttt{Action} requirement confirmation must be defined and included  \\ \hline
        C5 & A receipt confirmation must be sent of required \texttt{actions}  \\ \hline
        \textbf{Requirements} &          \\ \hline
        R1 & A risk boundary should be set to avoid the increase or decrease of water level in the ballast tank to a dangerous level. \\ \hline
        R2 & A risk boundary should be set to define and confirm channel integrity. \\ \hline
    \end{tabular}
\end{table}


\section{CyberShip Analysis using STRIDE}
\label{sec:stride_cybership}

\subsection{Methodology}
\label{subsec:methodology}

STRIDE stands for Spoofing, Tampering, Repudiation, Information Disclosure, Denial of Service and Elevation of Privilege.
The method was developed by Loren Kohnfelder and Praerit Garg in 1999~\cite{risk_stride}.
Spoofing is the ability of the attacker to pretend as someone or something else.
Generally, spoofing violates the authentication property of the system.
The modification or disruption of network or data of the system is known as tampering.
It violates the integrity of the system.
Repudiation is a threat that refers to someone's allegation that didn't perform an action that impacts the system's operation.
It violates the non-repudiation of the system.
Information disclosure is a threat which discloses confidential information to the people who are not suppose to have access to it.
It violates the confidentiality of the system.
The threat of Denial of Service (DoS) disrupts the availability of the system by consuming the resources required for the system to operate.
STRIDE model helps in identifying the potential threats and vulnerabilities during the design phase of the system.

The risk analysis is carried out by considering the likelihood of the threat and its impact. For the risk analysis, we used the criteria defined in Table~\ref{tabl:impact_criteria} and Table \ref{tabl:likelihhod_criteria}.

\subsection{STRIDE Application on CyberShip}
\label{sec:criteria_evaluation}

In this Section, STRIDE analysis on the CyberShip framework is presented, considering components and their interaction with each other.
It helps to get results which is valid regardless of its deployment in the framework.
Due to space constraints, STRIDE is applied on critical components of the CyberShip framework.
'I', 'L' and 'R' denotes impact, likelihood and risk in the Table \ref{tabl:engine_controller},Table \ref{tabl:ballast_controller} and Table \ref{tabl:IBC_controller}.

\begin{figure}[h]
    \includegraphics[width=0.9\textwidth]{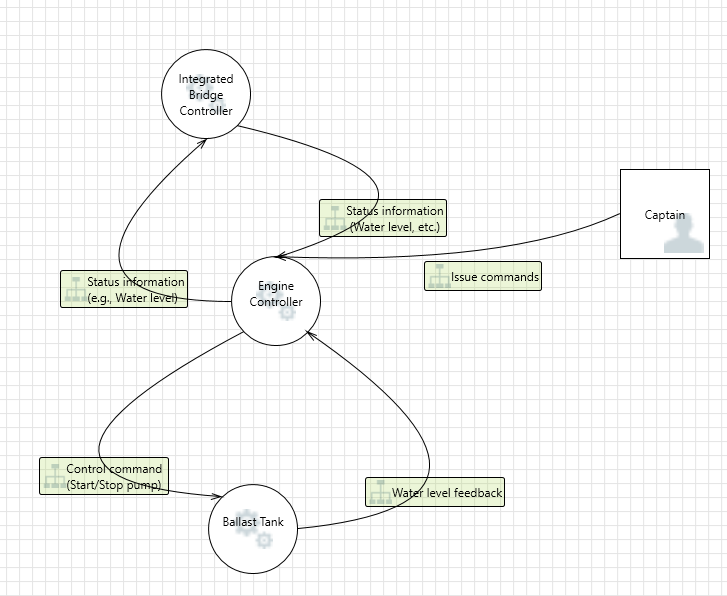}
    \centering
    \caption{Asset Diagram for CyberShip}
    \label{fig:assets_stride}
\end{figure}

Fig.~\ref{fig:assets_stride} shows components of the system with data flows for the STRIDE analysis.
Integrated bridge controller, engine controller and ballast tank are considered for the analysis.
It analyzes threats against each component that could be exploited by an adversary to compromise the whole system.
There are two ways to perform STRIDE analysis: 1) STRIDE-per-element; and 2) STRIDE-per-interaction.
STRIDE-per-element focuses on a set of threats against each components.
It makes it easier to enumerate threats.
However, many a times threats come up because of interaction among the components.
STRIDE-per-interaction finds threats against interaction between the components.

As we can see, tables show the high, medium and low level threats per components.
It can help in designing effective mitigation for each threats according to the different components requirements and risk levels.

\begin{table}[]
    \caption{Impact Criteria}
    \label{tabl:impact_criteria}
    \begin{tabular}{|l|l|}
        \hline
        High   & \begin{tabular}[c]{@{}l@{}}1. Loss of propulsion and endangering life of  crew members.\\ 2. It can damage the system.\\ 3. It can result in the financial and customer loss.\\ 4. It can cause system malfunction.\\ 5. It can affect the availability of the system.\end{tabular} \\ \hline
        Medium & \begin{tabular}[c]{@{}l@{}}1. It can impact the integrity of the system.\\ 2. It can cause information disclosure.\\ 3. It can cause procedure disruption in real time.\end{tabular}                                                                                                \\ \hline
        Low    & \begin{tabular}[c]{@{}l@{}}1. It can cause the operational delay in non-critical procedures.\\ 2. It may cause information disclosure of non-sensitive data.\end{tabular}                                                                                                           \\ \hline
    \end{tabular}
\end{table}

\begin{table}[]
    \caption{Likelihood Criteria}
    \label{tabl:likelihhod_criteria}
    \begin{tabular}{|l|l|}
        \hline
        Very Likely & \begin{tabular}[c]{@{}l@{}}1. The adversary is highly motivated and capable, and there are no deployed countermeasures.\\ 2. Existing popular exploits which can be executed remotely.\\ 3. System is directly exposed to the Internet. \\ 4. It can damage the system and provide benefit to attacker. \\ 5. Up to once in 6 months.\end{tabular}                                                                                                 \\ \hline
        Medium    & \begin{tabular}[c]{@{}l@{}}1.The adversary is highly motivated and capable, while the system countermeasures are not \\ enough to prevent the attack.\\ 2. The system's vulnerability is widely known, but it requires physical presence to launch the attack. \\ to exploit the vulnerability in the system.\\ 3. Systems are not directly exposed to the Internet.\\ 4. It can damage the system and provide some benefit to attacker.\\ 5. Multiple layers of systems need to be compromised. \\ 6. Once in 6 months up to once in 1 year.\end{tabular} \\ \hline
        Low         & \begin{tabular}[c]{@{}l@{}}1. The attacker is not highly motivated or does not have the necessary knowledge\\  to perform an attack, or deployed countermeasures are sufficient.\\ 2. It requires administrative rights to launch the attack.\\ 3. The system is not connected with external networks or systems. It requires physical presence. \\ 4. It can not damage the system and provide much benefit to attacker.\\ 5. Once in 2 years or less.\end{tabular}             \\ \hline
    \end{tabular}
\end{table}


\begin{table}[h]
    \caption{Engine Controller}
    \label{tabl:engine_controller}
    \begin{tabular}{|l|l|l|l|l|}
        \hline
        Threat (T)                 & Engine Controller                                                                                                                                                                                                                        & I & L & R \\ \hline
        Spoofing (S)               & \begin{tabular}[c]{@{}l@{}}Engine controller can be spoofed to the ballast tank pump and \\ to the IBC.  However, it requires physical presence or multiple layers\\ of systems need to be compromised.\end{tabular}                     & H & M & H \\ \hline
        Tampering (T)              & \begin{tabular}[c]{@{}l@{}}Control command from the EC to the ballast tank pump can be\\ tampered. It can cause damage to the ship.\\ But, physical presence is needed or multiple layers\\ of systems should be compromised.\end{tabular} & H & M & H \\ \hline
        Repudiation (R)            & \begin{tabular}[c]{@{}l@{}}EC can claim that it did not perform the command or received the data.\\ It can impact the root cause analysis of an incident.\end{tabular}                                                                   & L & L & L \\ \hline
        Information Disclosure (I) & \begin{tabular}[c]{@{}l@{}}It will not negatively impact the operation of the ship. Moreover, it\\ will not provide much benefit to attackers.\end{tabular}                                                                              & L & L & L \\ \hline
        Denial of Service (D)      & \begin{tabular}[c]{@{}l@{}}The availability of the controller is very important. It can cause\\ human safety issue.\end{tabular}                                                                                                         & H & M & H \\ \hline
        Elevation of Privilege (E) & \begin{tabular}[c]{@{}l@{}}If attacker gets administrative rights he can execute commands which\\ can damage the ship. But, it is difficult since the \\ engine controller is not directly exposed to the Internet.\end{tabular}                                                                                                      & H & L & M \\ \hline
    \end{tabular}
\end{table}

\begin{table}[h!]
    \caption{Ballast Tank}
    \label{tabl:ballast_controller}
    \begin{tabular}{|l|l|l|l|l|}
        \hline
        Threats                    & Ballast Tank Pump                                                                                                                                                                                                                                                     & I & L & R \\ \hline
        Spoofing (S)               & \begin{tabular}[c]{@{}l@{}}Computer system managing ballast tank can\\ be spoofed because of malware.  But, it \\ requires physical presence or multiple \\ systems such as engine controller, Integrated\\ Bridge Controller should be compromised.\end{tabular}     & H & L & M \\ \hline
        Tampering (T)              & \begin{tabular}[c]{@{}l@{}}Tampering with this system and data in transit can \\ cause critical damage and imbalance the ship.\\ But, it is difficult to tamper with the hardwired \\ signals from ballast tank.\end{tabular}                                         & H & L & M \\ \hline
        Repudiation (R)            & \begin{tabular}[c]{@{}l@{}}It can claim that it did not perform the command\\ or received the data.  But, it is very difficult as \\ every action happens due to some events.\end{tabular}                                                                            & L & L & L \\ \hline
        Information Disclosure (I) & \begin{tabular}[c]{@{}l@{}}Disclosure of  information will not negatively\\ impact the operation of the ship. Moreover, it\\ will not provide much benefit to attackers.\end{tabular}                                                                                 & L & L & L \\ \hline
        Denial of Service (D)      & \begin{tabular}[c]{@{}l@{}}The availability of the system is very important. \\ It can impact the operation of the ship.\end{tabular}                                                                                                                           & H & L & M \\ \hline
        Elevation of Privilege (E) & \begin{tabular}[c]{@{}l@{}}If attacker gets administrative rights he\\ can issue commands which can cause imbalance\\ and damage to ship. But, it is difficult to get the \\ administrative privileges as it is not directly \\ exposed to the Internet.\end{tabular} & H & L & M \\ \hline
    \end{tabular}
\end{table}

\begin{table}[t!]
    \caption{Integrated Bridge Controller}
    \label{tabl:IBC_controller}
    \begin{tabular}{|l|l|l|l|l|}
        \hline
        Threats                    & Integrated Bridge Controller (IBC)                                                                                                                                                                                                                                                         & I & L & R \\ \hline
        Spoofing (S)               & \begin{tabular}[c]{@{}l@{}}Spoofing can cause damage to the system and to \\ crew members. It can be due to malware spread\\ through online access or USB. The IBC's exposure to the \\ Internet is high.\end{tabular}                                                                     & H & M & H \\ \hline
        Tampering (T)              & \begin{tabular}[c]{@{}l@{}}Tampering with this system and data in transit can \\ cause operational problem to other connected components\\ and damage the ship.\end{tabular}                                                                                                               & H & M & H \\ \hline
        Repudiation (R)            & \begin{tabular}[c]{@{}l@{}}It can claim that it did not send the information\\ or received the data.  But, it is very difficult as \\ every action happens due to some events.\end{tabular}                                                                                                & H & L & M \\ \hline
        Information Disclosure (I) & \begin{tabular}[c]{@{}l@{}}Disclosure of  information will not negatively\\ impact the operation of the ship. Moreover, it\\ will not provide much benefit to attackers.\end{tabular}                                                                                                      & L & L & L \\ \hline
        Denial of Service (D)      & \begin{tabular}[c]{@{}l@{}}The availability of the system is very important. \\ It can negatively impact the operation of the ship.\end{tabular}                                                                                                                                           & H & M & H \\ \hline
        Elevation of Privilege (E) & \begin{tabular}[c]{@{}l@{}}If attacker gets administrative rights he\\ can forward wrong information which can cause\\ operational problem and damage to ship. \\ But, it is difficult to get the administrative \\ privileges as it is protected by the firewall.\end{tabular} & H & M & H \\ \hline
    \end{tabular}
\end{table}

Table~\ref{tabl:engine_controller}, Table~\ref{tabl:ballast_controller} and Table~\ref{tabl:IBC_controller} highlights STRIDE threats.
From the analysis we can see that for CyberShip spoofing, tampering and denial of service are the most critical threats.
In this analysis we assume that there are some countermeasures are deployed but they are not enough.
In addition to this, we also assume that the systems are not directly exposed to the Internet i.e. they are protected by firewalls and other systems.
However, Integrated Bridge Controller (IBC) has high exposure to the Internet as compared to the Engine Controller (EC) and computer system managing ballast tank.
Therefore, in the STRIDE analysis in all the cases likelihood of threat is taken as a medium.

As we can see in Table~\ref{tabl:engine_controller} spoofing, tampering and denial of service are the most critical threats for the engine controller that can directly damage the CyberShip.
If engine controller is spoofed by an adversary then command which is issued from it can be executed.
For example, if engine controller issues control commands (e.g., start or stop pump) then ballast tank can start pumping water.
Moreover, if the status level is tampered in this process by adversary then ballast tank can increase the water to a dangerous level in the tank which can imbalance the ship.

Table~\ref{tabl:ballast_controller} shows the STRIDE threat to ballast tank.
As we can see resultant risks are not very high in the case of ballast tank because it is not connected to the Internet and there are multiple layers of systems above it.
So, the likelihood of direct attack is low in the case of ballast tank.
Moreover, out of six STRIDE threats four can have high impact on the system.
Those four threats are spoofing, tampering, denial of Service (DoS) and Elevation of Privilege (EoP).
For example, tampering of feedback information from ballast tank to engine controller can damage the system as based on this information engine controller can issue command to either increase or decrease the water level in the ballast tank.
Moreover, computer system managing ballast tank can be spoofed and it can provide wrong information such as water level feedback to the engine controller that can damage the CyberShip.
Denial of service can happen on the computer system of ballast tank from the engine controller side once the engine controller is compromised.
It can also happen if the service engineer goes for some maintenance work and plug his infected USB in the computer system of the ballast tank.
Because of virus and malware spread through the infected USB can cause cyber attack on the computer system managing the ballast tank.
However, as computer system managing ballast tank is not directly connected to the Internet it won't have much impact.
However, if mis-configuration happens, and these systems get connected to the Internet then it can damage the CyberShip.
Similarly, elevation of privilege can only happen if the attacker is physically present or due to some mis-configuration it gets connected to the Internet and malware or viruses are present in the computer system of the ballast tank.

Table~\ref{tabl:IBC_controller} highlights the STRIDE threats to the IBC.
Spoofing, Tampering, Denial of service, and Elevation of privilege are the most critical for the IBC as it can impact the operation of the CyberShip.
For instance, if the information coming from the shore center is spoofed and tampered by an adversary then it can impact the operation of the CyberShip.
Moreover, in case navigational details provided by the IBC to the engine controller is tampered it can damage the CyberShip.
Because, CyberShip can rerouted or water level in the ballast tank can be increased or decreased.
To prevent this, we need to encrypt information in-transit and input validation at the engine controller and ballast tank to prevent the damage.

\section{CyberShip Analysis using CORAS}
\label{sec:coras_cybership}

CORAS is a model based risk analysis methodology~\cite{coras}.
It uses Unified Modelling Language (UML) for threat and risk modelling.
The UML is used to model the system under consideration and the context.
It also offers a tool to support documenting, maintaining and reporting risk analysis results.
CORAS is not only focused on identifying security requirements but is oriented towards performing holistic risk assessment.
The CORAS method is comprised of eight step process.
The authors in~\cite{coras} have described all the detailed stages of CORAS methodology in identifying the risks.
For brevity purpose, here the main focus is on identifying threats and how they can impact the availability of the CyberShip framework.

\begin{figure}[h]
    \includegraphics[width=0.9\textwidth]{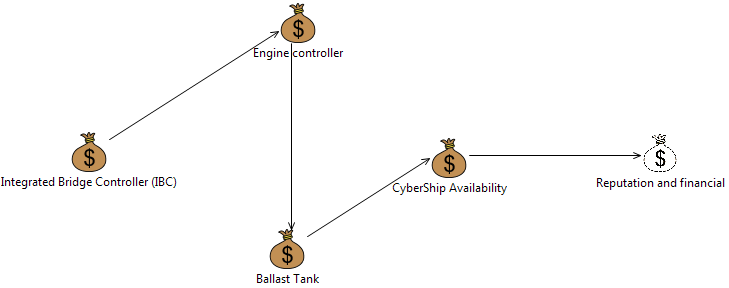}
    \centering
    \caption{Asset Diagram for CyberShip}
    \label{fig:assets}
\end{figure}

The first step in the CORAS methodology is to identify System under Consideration (SuC) or assets.
An asset could be anything such as hardware, software, people, reputation, etc.
Fig.~\ref{fig:assets} demonstrates the system under consideration for the threat identification and risk assessment.
It also shows how the different assets or components interact with each other on a high-level.
Three main assets or components which are selected for the risk assessment are: Integrated Bridge Controller, Engine Controller and Ballast tank.
The asset shown as CyberShip availability in Fig.~\ref{fig:assets} is directly impacted by the compromise of these assets.
Reputation of the organization or financial situation is indirectly affected due the impact on the availability of the CyberShip.

The main step after asset identification is risk identification using threat diagrams.
Threat diagrams represent possible scenarios which must be considered threats.
The purpose of this stage is to identify threats for the CyberShip framework.
We have considered threats because of adversary/hacker, crew, and component failure.
Generally, threats try to exploit the vulnerabilities in the system to harm the system.
In the analysis, we mainly considered threats initiated by adversary (deliberately), crew (accidental), and component failure.

\begin{figure}[h]
    \includegraphics[width=0.9\textwidth]{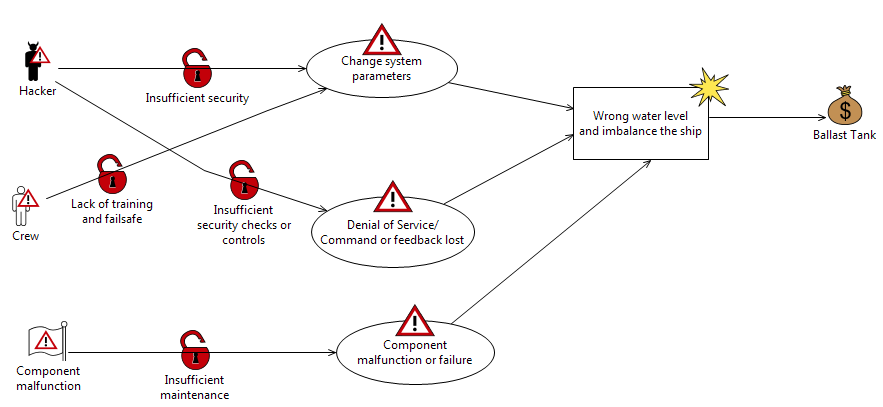}
    \centering
    \caption{Threat Diagram for Ballast Control System}
    \label{fig:threat_bcs}
\end{figure}

As we can see in Fig.~\ref{fig:threat_bcs}, unwanted incidents that can damage the ballast tank is wrong water level, it can imbalance the ship.
Change in system parameters, Denial of Service (DoS) and component or system malfunction can cause the wrong water level which can damage in the ballast tank.
For example, because of the DoS attack, ballast tank might not receive the command from the engine controller or feedback to the engine controller from the ballast tank might be lost that can lead to wrong water level that can imbalance the ship.
Moreover, lack of proper maintenance can cause system malfunction or component failure due to which ballast tank cannot receive or operate on the commands which can also lead to unwanted incidents.

\begin{figure}[h]
    \includegraphics[width=0.9\textwidth]{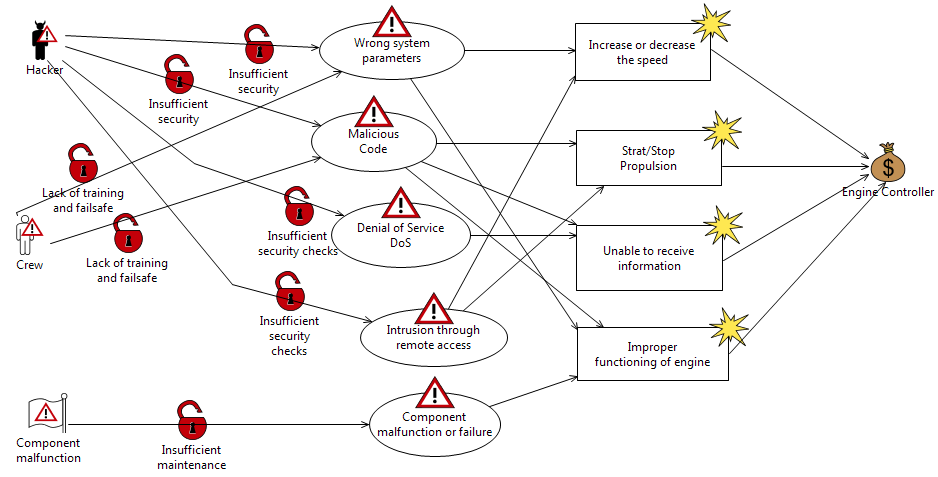}
    \centering
    \caption{Threat Diagram of Engine Controller}
    \label{fig:threat_engine}
\end{figure}


As shown in Fig.~\ref{fig:threat_engine}, threat scenarios comprising of wrong system parameters and component failure can lead to the compromise of integrity and availability of the system.
This in result can damage the engine controller.
Adversary can get the advantage of the lack of sufficient security measures and input validation to provide wrong system parameters to damage the engine controller.
For instance, adversary can input wrong parameters due to which engine can consume more fuel or get heated up which can damage the engine.
Parameters can be changed to a wrong value by insufficiently trained crew members and service engineers which can also damage the engine.
Wrong parameters can cause improper functioning of the engine controller that can damage it.
Moreover, lack of proper maintenance can also cause the component failure which can cause safety and security issues.

\begin{figure}[h]
    \includegraphics[width=0.9\textwidth]{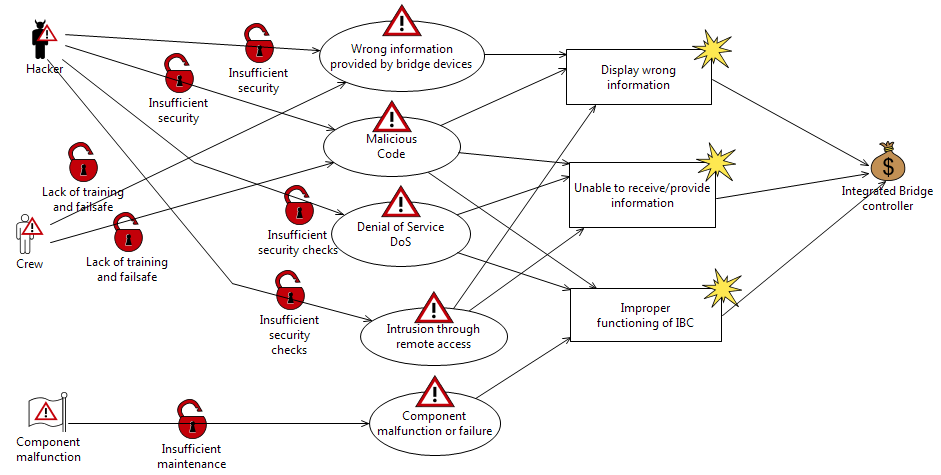}
    \centering
    \caption{Threat Diagram for Integrated Bridge Controller}
    \label{fig:threat_IBC}
\end{figure}

Fig.~\ref{fig:threat_IBC}, shows the threat scenarios which can lead to compromise of the Integrated Bridge Controller.
Many bridge devices (e.g. AIS, ECDIS) are vulnerable to cyber attacks because of lack of authenticity and integrity checks.
For example, due to the compromise of ECDIS, IBC will display wrong route information.
Because of the lack of application whitelisting and anti-virus protection ECDIS is vulnerable to malware attacks.
It can be exploited by adversaries to feed wrong information to the IBC, which can result in displaying wrong information.
It can result in the compromise of integrity and availability of the IBC.

\begin{figure}[h]
    \includegraphics[width=0.9\textwidth]{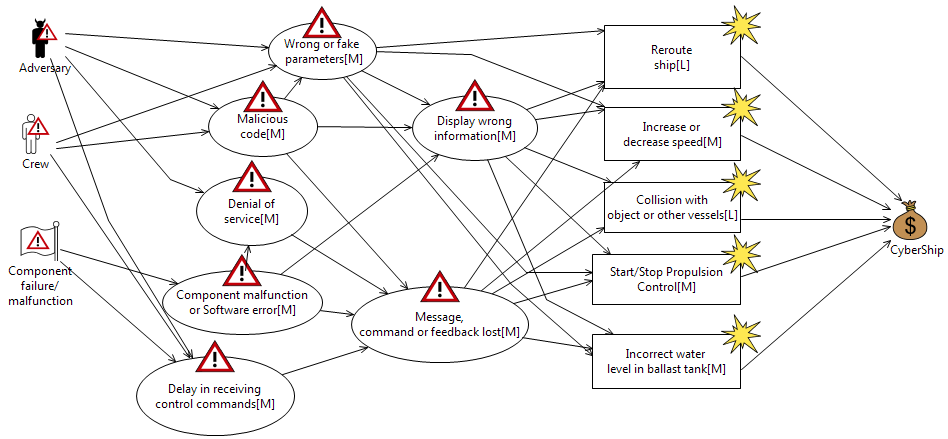}
    \centering
    \caption{Threat Diagram for CyberShip}
    \label{fig:threat_CyberShip}
\end{figure}

Fig.~\ref{fig:threat_CyberShip} illustrates different threat scenarios and likelihood which can damage the CyberShip.
Here, we assume that there is not much security is implemented.
More specifically, we say that it is initial level risk assessment to identify different threat scenarios and unwanted incidents which can damage CyberShip.
So, in most cases the likelihood of threat scenarios are assigned as "medium".
It can be seen in the Fig.~\ref{fig:threat_CyberShip} that wrong or fake parameters can lead to a threat scenario of displaying wrong information.
As a result of this, unwanted incidents like reroute of the ship, increase/decrease of speed, and start or stop of propulsion control can occur.
For instance, wrong or fake parameters provided to the engine controller can result in the increase or decrease of speed of the ship.
Similarly, IBC can display wrong information because of malicious code which can result into rerouting of the ship.
Denial of Service (DoS) attack can lead to the loss of control commands and feedback messages which can damage CyberShip.
These unwanted incidents can damage the CyberShip and lead to the compromise of its availability.
Incidents such as reroute and collision of the ship is assigned "low" likelihood because crew or captain of the ship keep monitoring the ship and surrounding through the window.
Moreover, these incidents can be easily noticed by crew.
Incidents such as "increase or decrease of speed" and "increase/decrease of water level in ballast" can get unnoticed for sometimes so "medium" likelihood is assigned.
Start/Stop propulsion control incident can occur quickly if engine controller is compromised so it is assigned the likelihood of "medium".

\begin{figure}[h]
    \includegraphics[width=0.9\textwidth]{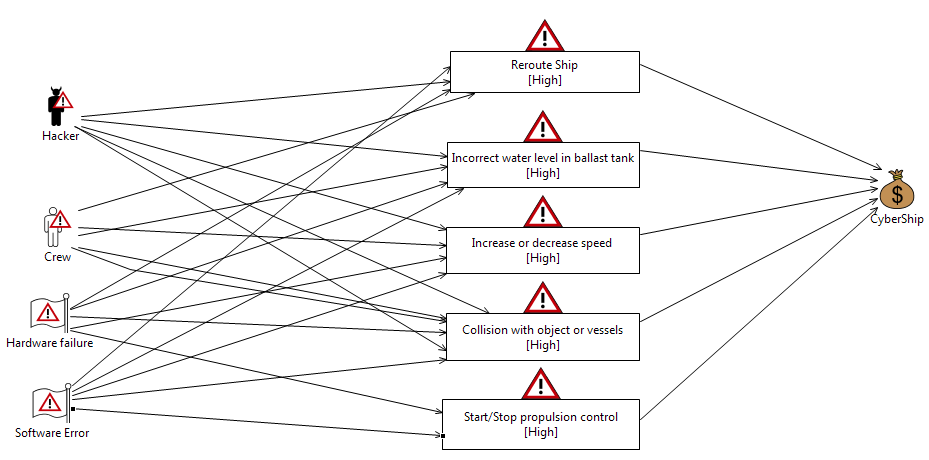}
    \centering
    \caption{Risk Diagram for Propagating Risk in CyberShip}
    \label{fig:cybership_risk}
\end{figure}

Fig.~\ref{fig:cybership_risk} shows different risk scenarios impacting the CyberShip which can result in the compromise of its availability.
Wrong water-level in the ballast tank can imbalance the ship which can result in accidents.
Similarly, increase or decrease of speed because of wrong parameters can damage the CyberShip's availability.
Since, it can cause congestion on the particular route and delay the arrival of the ship at the port.

\begin{figure}[h!]
    \includegraphics[width=0.9\textwidth]{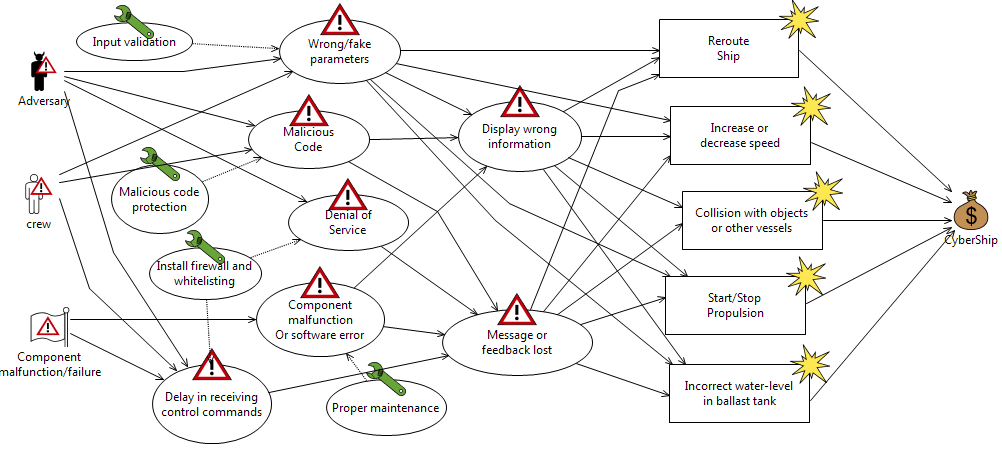}
    \centering
    \caption{Threat Mitigation Diagram for CyberShip}
    \label{fig:threat_mitigation}
\end{figure}

Fig.~\ref{fig:threat_mitigation} illustrates threat mitigation methods which can be deployed to mitigate the threats that can damage CyberShip.
The threat of wrong or fake parameters can be minimized with the deployment of \texttt{input validation}.
This will prevent changing the parameters in the controllers to beyond a threshold value.
Denial of Service (DoS) threat can be mitigated with whitelisting and deployment of proper firewall rules.
Because of brevity purpose only few mitigation mechanisms are shown in Fig.~\ref{fig:threat_mitigation}.
It is to be noted that other mitigation mechanisms can be deployed as well.
For example, to mitigate the threat of malicious code, authenticity and integrity checks can also be deployed along with malicious code protection.
To protect from component malfunction or software error, fail-safe and error handling should also be in place along with timely and proper maintenance.
Moreover, if system fails it should not disclose the sensitive information.

As we can see CORAS helps in identifying the threats, vulnerabilities and risks in a structured way through threat and risk diagrams.
It helps in visualizing threats and risks and analyse the way it can propagate and damage the system.
In the next Section, we compare three risk assessment methodologies that are STRIDE, STPA and CORAS.

\section{Analysis comparison between STPA, STRIDE and CORAS}
\label{sec:risk_analyses_comparison}

Risk assessment methodologies have some common stages in identifying and estimating threats and risks.
Some common stages in risk and threat assessment methodologies are:
\begin{itemize}
    \item \textbf{Establish the System under Consideration and context:} It is important to identify system under consideration, scope, likelihood and impact criteria before delving into the risk assessment.

    \item \textbf{Model of the System:} An explicit model of the system helps in identifying the threats, vulnerabilities, impact and resultant risk without getting bogged down in too many details.
    There are many different types of diagram that are helpful to model the system.
    For instance, data flow diagrams (DFDs), UML, and state diagrams are frequently used to design the model of the system.
    The main purpose of these diagrams is to present the high-level architecture of the system along with the information flow so that all the stakeholders involved in the risk assessment has the same understanding.
    Among these diagrams DFDs are the most frequently used diagram.

    \item \textbf{Identify Threats or Risks:} It is important to identify what aspects of system can go wrong and what are possible threats and hazards.
    There are many methods to finding threats and hazards.
    Generally, these are identified by organizing a workshop with system's expert.

    \item \textbf{Risk Estimation:} Risk estimation is to understand the nature of threat or risk and to find out its likelihood.

    \item \textbf{Risk Evaluation:} In this stage, risk estimation results are compared with the pre-defined criteria to determine whether the risk is acceptable or tolerable.

    \item \textbf{Risk Mitigation:} It is about planning and implementing countermeasures to mitigate or minimize the impact of risk.
\end{itemize}

\subsection{Analysis of STPA, STRIDE and CORAS}
\label{sec:comparison_evaluation}

In this Section, a comparison between STPA, STRIDE and CORAS is presented.
The aim is to highlight the key features of these methodologies and analyze the context in which they can perform better than the others.
Table~\ref{tabl:STPA_STRIDE_CORAS} highlights some key stages of STPA-Sec, STRIDE and CORAS methodologies.

The detailed control structure of the system is established for the hazard analysis using STPA-Sec methodology.
Table~\ref{tabl:UCA} shows unsafe control actions or events that could cause hazardous state.
As we can see, 17 unsafe control actions or commands have been identified in the framework using STPA-Sec.
It is evident that one hazard can be triggered by more than one unsafe control actions.
For example, uncontrolled maneuvering of the ship can be due to \textbf{UCA1.1},\textbf{UCA1.2}, \textbf{UCA1.11}, \textbf{UCA1.12}.

Hazards identified through STPA-Sec method can be put into five following categories as suggested by ~\cite{leveson2011engineering}:
\begin{itemize}
    \item Component failure
    \item Component interaction
    \item Software fault
    \item Human error
    \item System error
\end{itemize}

These five categories cover all the system components.
Nowadays, cyber-physical-systems (CPS) are comprised of software. hardware, human and interacting components.
Therefore, these five categories cover all the components of the CyberShip framework.
It should be noted that the STPA-Sec method identifies unsecure control actions along with problems caused because of design errors, software faults, component interaction, human in the loop, and cyber attacks.
Moreover, it also identifies security constraints that can be enforced to prevent system from entering into vulnerable state which leads to damage.
The main aim of STPA-Sec is to identify unsafe and unsecure control actions that can cause damage to the system.
More specifically, it identifies hazards or conditions that can lead the control actions to unsafe or unsecure state which can compromise or damage the system.
However, many well known cyber threats and vulnerabilities might not be covered by this.
For example, as shown in the Table~\ref{tabl:UCA} unsafe control actions such as \textbf{UCA1.1}, \textbf{UCA1.2}, \textbf{UCA1.4} and \textbf{UCA1.5} can occur because \texttt{Engine Controller} is compromised.
However, threats and vulnerabilities that can compromise the \texttt{Engine controller} can not be identified at this stage.
A detailed threat modelling or cyber risk analysis at the component level is required in identifying threats and vulnerabilities that can lead to the compromise of \texttt{Engine controller}.
It is important to include security-related accidents into definition of accidents while performing STPA-Sec.
STPA-Sec uses a set of causal factors for security.
However, in this causal factors focus is on integrity and availability, confidentiality is not considered.
Moreover, security related losses and hazards that can cause these losses must be considered in detailed to integrate security requirements in the STPA-Sec analysis.
Furthermore, STPA-Sec does not employ a threat model to consider new causal factors.


It is clear from the STRIDE analysis that it offers a systematic approach to identify cyber threats against each component of the system.
It analyzes cyber threats corresponding to security properties such as authentication, authorization, system integrity, confidentiality, repudiation and availability.
As shown in the analysis of STRIDE, the impact, likelihood and the resultant risks are described in terms of qualitative values and very much subjective.
As we can see the STRIDE analysis helps in identifying different types of cyber threats.
The STRIDE methodology is very good in enumerating the list of cyber threats which can compromise the system.
However, it is important to identify safety as well as security threats in the cyber physical systems (CPS) such as CyberShip.
Therefore, cyber threats identified through STRIDE mechanism can be used as an input to other mechanisms such as CORAS and STPA-Sec.
It can assist in identifying more threats in a structured way.

CORAS is also a top-down and asset-based approach of identifying risks.
It relies on asset, threat, risk and treatment diagrams to identify threats, vulnerabilities, and risks.
This method is very helpful in visualizing the way threat can propagate and lead to unwanted incidents that can damage the system.
Generally, CORAS method identifies threats in a brainstorming session.
As shown in the CORAS analysis that it can also identify safety threats and risks such as component malfunction or software error and the way these threats can lead to incidents that can damage the system.
Moreover, it helps in identifying the nature of vulnerabilities that can cause threats as compared to STPA-Sec and STRIDE approach.
However, CORAS lacks the structured process of finding out threats that can compromise the system.
It relies on the expert of the system to identify threats and vulnerabilities and depends on experience.
However, threats identified through STRIDE method can be used as input while doing CORAS analysis, it can make CORAS more structured and will help in identifying more threats and vulnerabilities which are not easy to identify.

\begin{table}[h!]
\caption{Comparison between STPA-Sec, STRIDE and CORAS Methodologies}
    \label{tabl:STPA_STRIDE_CORAS}
\begin{tabular}{|l|l|l|l|}

\hline
Stages & STPA-Sec & STRIDE & CORAS \\
\hline
\begin{tabular}[c]{@{}l@{}}System Under\\ Consideration \\ and Context\end{tabular} & \begin{tabular}[c]{@{}l@{}}System objective: Identify \\ the goal for which  the\\ system is designed\\ and system boundaries\end{tabular} &
\begin{tabular}[c]{@{}l@{}}System is identified on\\ which analysis has to \\ be performed.\end{tabular} &
\begin{tabular}[c]{@{}l@{}}Asset identification is\\  the first step\\ in CORAS.\end{tabular}\\
\hline
Model of the System &
\begin{tabular}[c]{@{}l@{}}Identify System Structure: \\ list\\ the controls, process models,\\ processes and operators \\ and their connections, \\ including hierarchy.\end{tabular} & \begin{tabular}[c]{@{}l@{}}System model is created \\ using DFD.\end{tabular} &
\begin{tabular}[c]{@{}l@{}}Model of the system\\ is created using\\ threat diagram.\end{tabular} \\
\hline
Identify Risks &
\begin{tabular}[c]{@{}l@{}}Identify Hazards: list \\ system state or conditions\\ that with\\ environmental \\ worst-case scenario, lead to\\ an unacceptable loss\\ \\ Identify requirements and\\ constraints: list the \\ controls by presence (passive) \\ and the controls by\\  action (active)\\ through detection, \\ measurement, diagnosis\\  or response.\end{tabular} & \begin{tabular}[c]{@{}l@{}}Identify threats: \\ spoofing,\\ tampering, repudiation, \\ information disclosure, \\ elevation of privilege,\\ DoS are identified.\end{tabular} & \begin{tabular}[c]{@{}l@{}}Threats, vulnerabilities \\ and unwanted incidents\\ are identified using\\ threat \\ diagrams.\end{tabular}                 \\ \hline
Risk Estimation &
\begin{tabular}[c]{@{}l@{}}Describe UCA:\\ list conditions when each\\ control action\\ or their lack creates a hazard. \\ List hazards through degradation\\ over time.\end{tabular} &
\begin{tabular}[c]{@{}l@{}}Determine likelihood and\\ impact in a brainstorming\\ session.\end{tabular} &
\begin{tabular}[c]{@{}l@{}}Risk diagram helps\\ in identifying and\\ estimating\\  risks.\end{tabular} \\
\hline
Risk Evaluation                                                                     & \begin{tabular}[c]{@{}l@{}}Evaluate UCA: \\ list causal scenarios \\ and additional\\ constraints from \\ the UCA analysis\end{tabular}                                                                                                                                                                                                                                           & \begin{tabular}[c]{@{}l@{}}Tolerable and \\ unacceptable\\ risks are identified \\ which need further\\ investigation.\end{tabular}                                          & \begin{tabular}[c]{@{}l@{}}Tolerable and \\ unacceptable\\ risks can be identified\\ through risk diagram.\end{tabular}                                 \\ \hline
Risk Mitigation                                                                     & \begin{tabular}[c]{@{}l@{}}List the additional design\\ requirements to implement\\ the additional constraints\end{tabular}                                                                                                                                                                                                                                                       & \begin{tabular}[c]{@{}l@{}}List countermeasures to \\ mitigate the threats.\end{tabular}                                                                                     & \begin{tabular}[c]{@{}l@{}}Through threat\\ mitigation diagram\\ countermeasures\\ are shown to mitigate \\ threats and\\ vulnerabilities.\end{tabular} \\ \hline
\end{tabular}
\end{table}

\section{Discussion}
\label{sec:discussion}

STPA-Sec uncovers more hazardous situations at the design level.
Also, by focusing the analysis on the system structure, STPA-Sec analysis results in design recommendations to secure shipping system against cyber attacks which are independent of the source of the attacks.
The main challenge with STPA-Sec is finding a method for identify cyber threats both at the component level and from mis-interactions between the components, which is consistent, thorough and not overly dependent on experience.
However, STPA-Sec can be extended with the STRIDE threats to identify new loss scenarios and requirements because of the lack of security.

CORAS offers a model based framework for identifying threats, vulnerabilities and risks.
Threat and risk diagrams are helpful in visualizing the way threats and risks can compromise and damage the CyberShip.
Countermeasures to mitigate threats are visualized through the threat mitigation diagram.
Threat mitigation diagram can help in eliciting security requirements for the system under consideration.
Moreover, safety risks can also be identified using CORAS approach.
But, the challenge in CORAS methodology is to find or enumerate different threats that can exploit vulnerabilities which can damage the system.
Because of the complexity of the system, it is very common to miss threat scenarios.
However, STRIDE threats can be considered during the CORAS based risk analysis.
Moreover, these threats can be mapped to confidentiality, integrity and availability features, i.e. threats which can compromise the confidentiality, availability and integrity of the system.
It can assist in performing risk assessment in a structured way by considering the main objective of the system.
For instance, availability is the major concern in the case of CyberShip.

STRIDE approach is very good at enumerating threat lists.
The disadvantage with STRIDE model is that it does not consider safety risks which is a major concern in the cyber physical systems such as CyberShip.
STRIDE method is very good for identifying computer security threats, however, it does not consider the physical aspects which are very important in cyber physical systems.
However, the list of threats found using STRIDE can be used in CORAS and STPA-Sec analysis to perform a  detailed risk assessment.
Furthermore, identification of STRIDE threats depends on the experience of the person facilitating risk assessment.
For instance, it is not easy to directly identify how spoofing can happen in the case of engine controller.
Below we use an example of using STRIDE threats as an input in the threat diagram of CORAS framework.

\begin{figure}[h!]
    \includegraphics[width=0.9\textwidth]{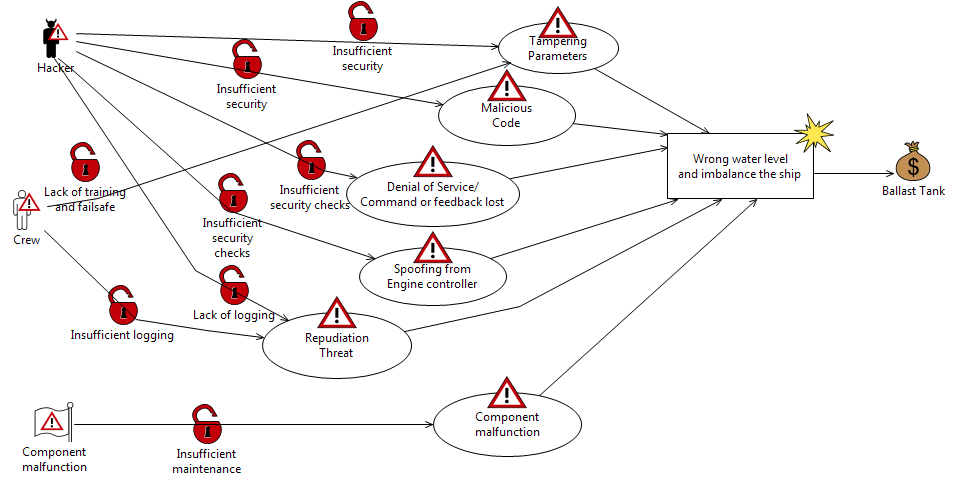}
    \centering
    \caption{Threat Diagram of Ballast Tank Considering STRIDE Threats}
    \label{fig:STRIDE_CORAS}
\end{figure}

As we can see in Fig.~\ref{fig:STRIDE_CORAS}, STRIDE threats such as tampering, DoS, spoofing and repudiation all these threats targeting ballast tank can lead to unwanted incident such as wrong water level in the ballast tank that can imbalance the ship.
An adversary can spoof as an engine controller and issue control commands to either increase or decrease the water level in the ballast tank to a wrong level.
It can happen because of insufficient security checks, lack of proper firewall rules \& segmentation, lack of input validation and application whitelisting, etc.
In this way, by using STRIDE threats during STPA-Sec and CORAS analysis, we can identify more threat scenarios in a structured way.

However, all three methodologies do not consider exploitability while performing risk assessment.
Exploitability specifies how easy it is to target the particular component in the system.
Moreover, during risk assessment it is also recommended to consider the level of attackers i.e. whether they are script kiddie or skilled attackers which is not considered during STPA-Sec, STRIDE and CORAS assessment.
It depends on the facilitator to include these criteria during risk assessment.




\section{Conclusion and future work}
\label{sec:conclusion}

In this article, we applied STPA-Sec, STRIDE and CORAS methodologies on CyberShip framework to perform risk assessment.
By applying these three methodologies, we identified the threat and hazard scenarios along with security requirements for the CyberShip framework.
In the use case, we focused the analysis on safety and security; but other concerns such as privacy can be considered.
The analysis helped in identifying the pros and cons of these methodologies when applied in the same framework.
We found that STPA-Sec is good at identifying hazards by looking at the control actions and the structure of the system.
CORAS provides a good framework for visualizing and identifying threats and unwanted incident scenarios.
However, it relies on the expert of the system in identifying that.
It would be difficult for somebody who is not an expert in those systems.
STRIDE offers a structured approach for highlighting threats to the system which can be coupled with CORAS or STRIDE to make them more effective.
Even, STRIDE can benefit from the control structure provided by STPA-Sec and threat mitigation diagram provided by CORAS in identifying risks.
We also extended CORAS method with STRIDE approach to identify more threats CyberShip frameworks.
As a future work, we intend to extend these methodologies in detail to identify the threats, risks and security requirements of CyberShip.

\section{Acknowledgement}
\label{sec:acknowledgement}

    The authors would like to acknowledge the funding provided by the Orients Fund by the Danish Maritime Fund (DMF) to the project CyberShip, "Cyber resilience for the shipping industry" at the Technical University of Denmark, DTU, for the period 2017-2020.


    \bibliographystyle{plain}
    \small
    \bibliography{references}

\end{document}